\documentstyle[prb,aps,twocolumn]{revtex}

\input epsf

\def\prb{Phys. Rev. B}
\def\prl{Phys. Rev. Lett.}

\def\be{\begin{equation}}
\def\ee{\end{equation}}
\def\ba{\begin{eqnarray}}
\def\ea{\end{eqnarray}}

\def\LCO{La$_{2}$CuO$_{4+\delta}$}
\def\LSCO{La$_{1.86}$Sr$_{0.14}$CuO$_4$}

\def\YBCO{YBa$_2$Cu$_3$O$_{7-\delta}$}
\def\BSCCO{Bi$_2$Sr$_2$CaCu$_2$O$_{8+\delta}$}

\def\C60{A$_x$C$_{60}$}
\def\LNSCO{La$_{1.6-x}$Nd$_{0.4}$Sr$_x$CuO$_{4}$}

\def\hts{high temperature superconductors}

\begin{document}

\twocolumn[\hsize\textwidth\columnwidth\hsize\csname@twocolumnfalse\endcsname

\title
{Stripe Liquid, Crystal, and Glass Phases of Doped Antiferromagnets}

\author{S.~A.~Kivelson$^1$ and V.~J.~Emery$^2$  }
\address
{1)  Dept. of Physics,
U.C.L.A.,
Los Angeles, CA  90095}
\address{
2) Dept. of Physics,
Brookhaven National Laboratory,
Upton, NY  11973-5000}
\date{\today}
\maketitle 

\begin{abstract}
A largely descriptive survey is given of the ordered 
phases of doped antiferromagnets, and of the long 
wavelength properties 
that can be derived from an order-parameter theory.  In particular, 
we show that the competition between the long-range Coulomb repulsion 
and the strong short-distance tendency of doped holes to coalesce into 
regions of supressed antiferromagnetism leads to a 
variety of self-organized charge structures on intermediate length 
scales, of which ``stripes'' are the most common, both theoretically 
and experimentally.  These structures lead to 
a rich assortment of novel electronic phases and crossover 
phenomena, as indicated in the title.  We use the high temperature 
superconductors as the experimentally best-studied examples of doped 
antiferromagnets.
\

\

\end{abstract}

]

Highly correlated materials have intermediate electron densities
and are frequently doped Mott insulators, 
so that neither the kinetic energy nor the potential 
energy is totally dominant, and both must be treated on equal 
footing.  The question arises, are there actual ``intermediate'' 
low temperature phases of matter which interpolate
between the high density ``gas'' phase (usually called a Fermi liquid) 
and the low density strongly insulating Wigner crystal phase?  We have 
shown that, at least in the case of lightly-doped 
antiferromagnets, the tendency of the 
antiferromagnet to expel holes always \cite{ekl,marder,manousakis,pryadko} 
leads to phase separation which, when frustrated by the long-range piece of 
the Coulomb interaction, leads \cite{topo,physicaC,ute} 
to the formation of states which are 
inhomogeneous on intermediate length scales and (possibly) time scales.
The  most common self-organized structures which result 
from these competing interactions\cite{ute,science} are ``stripes'', by which,
we generally mean $d-1$ dimensional antiphase domain walls across which the 
antiferromagnetic order changes sign, and along which the doped holes 
are concentrated;\cite{larged}  the term ``stripe'' is, of course, a 
reference to the important two-dimensional case relevant to the 
high temperature superconductors.

Even if the discussion is confined to ordered phases of doped 
antiferromagnets we are left with an exceedingly complex problem.  
Many varieties of order have been observed in doped antiferromagnets, 
including spin and charge order and, of course, superconductivity.  
The spin and charge order can be commensurate or incommensurate, and 
both can be ideal or glassy. There are also various structural phases, 
such as the tetragonal and orthorhombic phases of {\LSCO}, which
may reflect important changes in the electronic state, as the structural 
order can couple to various forms of ``electronic liquid crystalline''
order. \cite{fradkin} Of course all of these types of order can compete or 
coexist in various ways.  

\section{Landau Theory of coupled CDW and SDW order}
We begin by discussing density-wave order, and in particular the 
interplay between spin-density wave (SDW) and charge-density wave 
(CDW) order.  
This can be analyzed most simply by studying the 
Landau theory of coupled order parameters.\cite{landau}  While it is 
possible to have various sorts of ``spiral'' spin phases, 
the only spin order in much of parameter space is collinear, so the 
discussion will be confined to this region.  The resulting phase 
diagram is shown schematically in Fig. 1.  Three features of the 
analysis, which is discussed in detail in Ref. 11, bear 
repeating:  1)  In order for the SDW order parameter $S_{\vec q}$ and the CDW 
order parameter $\rho_{-\vec Q}$ to couple in the 
lowest possible order (third), it is necessary that the ordering 
vectors satisfy the relation ${\vec Q}=2 {\vec q}$, or in other 
words, the wavelength of the SDW is twice that of the CDW;  this gives
precise meaning of the concept\cite{topo} of ``topological doping'' 
and implies that the charge is effectively concentrated along 
antiphase domain walls in the magnetic order.  2)  It is possible to 
have a phase with CDW order, but no SDW order, whereas SDW order always 
implies CDW order.  This is important to bear in mind when thinking 
about the experimentally determined phase diagram of the high 
temperature superconductors or any other doped antiferromagnet, since 
there are many good probes (such as NMR, $\mu$SR, and neutron scattering) 
that are sensitive to spin order or fluctuations, but fewer that are 
sensitive to charge order. Where incommensurate spin order is 
detected, we can directly infer the existence of charge order, but 
where no magnetic order is observed, there may or may not exist as yet 
undetected charge order.  3)
Although Landau theory by its very character is relatively insensitive 
to the microscopic considerations conventionally referred to as the 
``mechanism'' of ordering, an important classification of mechanisms 
follows directly from these considerations.  If, upon lowering 
temperature, CDW order is encountered first and SDW order is either
entirely absent or only appears at lower 
temperatures when the CDW order is already well developed, the 
density wave transition is ``charge-driven'', and we can 
infer that the SDW order is in some sense parasitic, {\it i.e.}
driven by the interaction with the CDW.  On the other hand, 
if both CDW and SDW order develop simultaneously, but with the
CDW order turning on more slowly at the transition according to 
$\rho_{-2\vec q}\sim \vec S_{\vec q}\cdot \vec S_{\vec q}$, then the 
ordering can be said to be ``spin driven''.  (Intermediate cases, in 
which the spin and charge ordering must be treated on an equal 
footing, are also possible.\cite{landau})  Hartree-Fock 
treatments\cite{zaanen} 
of stripes lead to spin-driven 
ordering while frustrated phase separation (and, indeed, experiments 
in the nickelates, manganates, and the appropriate cuprates) imply 
that the ordering is, in fact, charge driven.  

\begin{figure}
\vspace{.8cm}
\hspace{1.15 in}
\epsfxsize=3.3in
\epsfbox{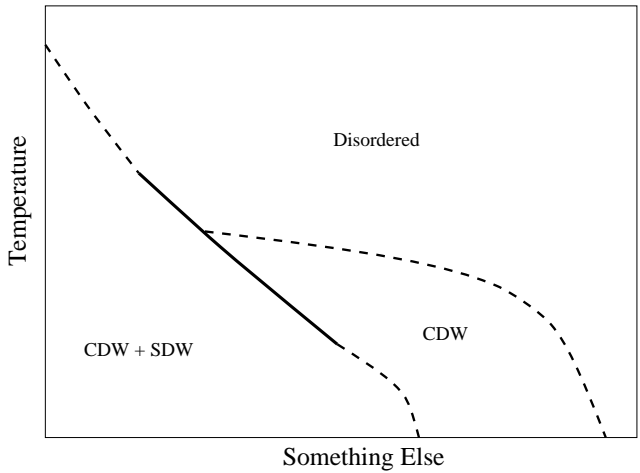}
\vspace{.8cm}
\caption{  Schematic phase diagram from the Landau theory of coupled 
CDW and SDW order parameters.  The y axis is temperature, and the x 
axis is a microscopic, material dependent parameter. 
}
\label{fig1}
\end{figure}

\section{ Concerning the Mechanism of Stripe Formation}

The charge ordering described above, which 
we think of as ordered arrangements of charged ``stripes'', differs 
from more usual CDW order in metals in that it is a 
consequence of frustrated phase separation, not
a Fermi-surface instability.  There are three important consequences: 
1)  Whereas conventional CDW order, as observed in many 
charge-transfer salts and bronzes, tends to oscillate in all directions,
the stripe order that we have in mind here involves a one dimensional 
modulation of the charge density; hence the name ``stripes''.  
2)  There can be additional low energy electronic degrees of freedom 
within a stripe, {\it i.e.} the stripe can be metallic and have its own  
spin dynamics. 3) The density of states at the Fermi level is increased,
\cite{markku} not decreased by the opening of a gap in the electron
energy spectrum, as for a CDW driven by a Fermi-surface instability.

The second point is exceedingly important in 
distinguishing mechanisms of stripe formation, and is a unique feature 
of stripes produced by competing long and short-range 
interactions.\cite{topo}  In any model with only short-range 
interactions, stripes, if they form at all, have a preferred hole 
density which is usually commensurate and, in turn, determines the stripe 
concentration. Consequently, in general, even if there were a region of 
parameters in which a putative 
stripe phase were not unstable to phase separation, it always would be 
insulating\cite{zaanen,white} with a substantial charge gap.
By contrast, 
whenever the stripes arise as a consequence of a competition between a 
long-range Coulomb interaction, and a short-range tendency to phase 
separation, the stripe concentration is a compromise between these two forces, 
so gaps in the intra-stripe excitation spectrum are no longer the rule.\cite{yamada}

Finally, self-organized stripe structures are an intrinsic property of
a doped antiferromagnet.  While they certainly couple in 
interesting material-specific ways to  external inluences,
such as the lattic modulation in {\BSCCO} or the chains in {\YBCO},
\cite{bianconi} these extrinsic effects should be viewed as 
reflecting the intrinsic stripe physics, rather than causing it.

\section{ Electronic Liquid Crystal Phases}
 
Following the work in Ref. 10 
we analyze the case in which intra-stripe 
metallic degrees of freedom interact with the 
fluctuating ``geometry'' of the stripe arrangement, resulting in a set 
of new states of matter, which in analogy with classical liquid 
crystals, we have named ``electronic liquid crystals.''  

By analogy with classical liquid crystals, we can  
readily deduce the schematic phase diagram, as shown in Fig. 2.  Here, 
the x axis is a quantum parameter, $\hbar \bar \omega$,
related to the transverse zero-point 
energy of the stripes, {\it i.e.} it measures the extent of quantum 
fluctuations of the stripe order.  These phases, which can 
exist at either zero or finite temperature, can be classified as 
follows:

{\bf 1)  Stripe crystal phases:}  Here translational symmetry in the 
direction perpendicular to the stripes is broken because the stripes are 
ordered, and translational symmetry along the stripe is broken because the 
electrons along the stripe form a CDW which is phase locked between stripes.
This phase is conceptually similar to the CDW phases that occur in 
charge transfer salts and, in common with them, it is insulating. 
In the important, but special cases in which the stripes have 
no low energy internal degrees of freedom, ({\it e.g.} stripes that are full
of holes or electrons), the insulating state is achieved without breaking 
translational symmetry along the stripe direction. 
At zero temperature in two spatial dimensions, the stripe order 
is always pinned at a period commensurate with the host crystalline 
lattice, while at non-zero temperature it can be commensurate or 
incommensurate, but where it is incommensurate, the positional order 
is only quasi-long range.  

{\bf 2)  Electronic smectic phases:}  If the stripe order is maintained, 
but the intra-stripe excitations remain unpinned, one obtains an 
electronic smectic phase.  This phase posseses the same broken 
symmetries transverse to the stripe direction as the stripe crystal 
phases (with all the same dimension specific considerations mentioned 
above), but simultaneously exhibits liquid-like behavior associated 
with the motion of electrons within a stripe, and the tunnelling of 
electrons between stripes.  Depending on additional interactions in 
the problem, this phase can either remain an electron liquid 
down to zero temperture, or can become superconducting below a 
critical temperature, $T_{c}$.  

A major portion of the work in Ref. 10 
addressed the issue of how the 
transverse fluctuations of the stripes controls both the transition
between the stripe crystal and electronic smectic phases, and the 
magnitude of the superconducting $T_{c}$ within the smectic phase.  
Basically, in the presence of a spin gap, the intra-stripe electron 
gas is prone to large superconducting and CDW fluctuations.  In two 
spatial dimensions (where within a stripe the electrons form an
effective one 
dimensional electron gas), it is known that both these susceptibilies 
typically will diverge as $T \rightarrow 0$, with the CDW being the 
more divergent.  In higher dimensions, more complicated possibilities 
exist, but the physics is not qualitatively different.  Now, because 
the CDW order involves short-wavelength density oscillations along 
the stripe, the CDW ordering on neighboring stripes is readily 
dephased by transverse stripe fluctuations;  through this mechanism, 
increasing transverse stripe fluctuations (either quantum or thermal) 
can easily be shown to stabilize the smectic phase at the expense of 
the stripe crystal.  Conversely, pair tunnelling between neighboring 
stripes, and hence the transverse component of the superfluid density,
is enhanced by transverse stripe fluctuations.  To see this, we note 
that the local pair-tunneling matrix element, $J_{pair}$, depends 
exponentially on 
the local separation, $w$, between stripes
\be
J_{pair}\approx J_{0}\exp[-\alpha w].
\ee
To get an idea for the physics, 
we average this quantity over transverse stripe fluctuations, keeping 
terms up to second order in a cumulant expansion, with the result
\be
\bar J_{pair}\approx J_{0}\exp[-\alpha \bar w] \exp[\alpha^{2} 
{(\Delta w)^{2}}/2]
\label{eq:Jofw}
\ee
where $
{\Delta w}$ is the variance of $w$.  Clearly, for fixed mean 
spacing $\bar w$ between stripes, the pair tunnelling is a strongly 
{\it increasing} function of $ 
{\Delta w}$.

{\bf 3)  Electronic (Ising) Nematic:}  When the transverse stripe 
fluctuations get sufficiently violent, they will certainly lead to a 
liquid state, with full translational symmetry.  However, it is 
possible that the general orientation of the stripes can persist beyond the 
melting transition.  In this case the electronic phase is liquid-like and 
translationally invariant, but it breaks the discrete rotational symmetry of
the host crystal. The nematic phase in fact breaks this discrete symmetry.  
For instance, in a tetragonal 
crystal with a four-fold rotational symmetry, the nematic phase would 
be electronically orthorhombic, with only a two-fold rotational 
symmetry surviving.\cite{abanov}  Of course, there is bound to be some 
back coupling between electronic and lattice structure, so such a 
phase would also be accompanied by a locally orthorhombic distortion of 
the crystalline lattice.  Like the smectic, the nematic phase can 
either remain a normal liquid down to zero temperature, or become 
superconducting.  

We have sketched a representative superconducting 
phase boundary as a dashed line in Fig. 2.  The logic governing 
its shape is as follows: Superconductivity requires both pairing, which 
occurs below a temperature $T_{pair}\sim \Delta(0)/2$,
where $\Delta(0)$ is the zero temperature magnitude of the 
superconducting (spin) 
gap or pseudogap, and phase ordering, which
sets in below a temperature $T_{\theta}$ which is 
proportional to the zero temperature superfluid density.\cite{nature}
It is clear that the general trend found in the smectic, 
in which an increase in the transverse fluctuations increases the transverse 
phase stiffness, should apply in the nematic phase close to the smectic
phase boundary.  Thus, so long as $T_{\theta}$ determines $T_{c}$,
it will be an increasing function of $\hbar \bar \omega$.
Of course, ultimately the transverse fluctuations 
become so violent that even the local concept of a stripe ceases to 
be well defined;  if we take the view that stripes are an essential 
ingredient in the pairing mechanism\cite{ekz} (equivalently, in the spin-gap 
formation) then $T_{c}$ must first rise and then decrease with 
increasing flucutations.  

We have shown the peak in $T_{c}$ near the 
nematic to isotropic phase boundary, because
we imagine that this is where the 
stripes start to fall apart at
a local level.  The points marked $C_{j}$ are quantum critical
points.  As drawn, the superconducting
phase boundary crosses the nematic phase boundary at a
tetracritical point, $T_1$, but it is equally possible that the
superconducting phase boundary could end at a bicritical point (say, roughly
where $T_1$ appears in the figure), and that beyond this there is
a  (possibly weak) first order  phase boundary
marking the simultaneous onset of superconducting and nematic order.
Alternatively, the superconducting phase might, in some circumstances,
lie entirely inside the nematic phase, if the quantum critical points
$C_2$ and $C_3$ were exchanged, in which case there would be no
multicritical point analagous to $T_1$.

{\bf 4)  Isotropic stripe liquid:}  With sufficiently large 
transverse fluctuations, all symmetry is restored, so an 
isotropic liquid phase results.  However, if sufficient 
stripe order persists on a local level, the resulting isotropic stripe liquid 
may be quite different from a non-interacting electron gas.  Since 
there is no symmetry distinction, it is possible that the evolution 
from a stripe liquid to a Fermi gas involves a crossover, but often, 
as in a liquid gas transition, a first order transition separates two 
qualitative different but asymptotically similar phases;  for that 
reason, we have included such a first order line, ending in a critical 
point, in Fig. 2.

\begin{figure}
\vspace{.8cm}
\hspace{1.15 in}
\epsfxsize=3.3in
\epsfbox{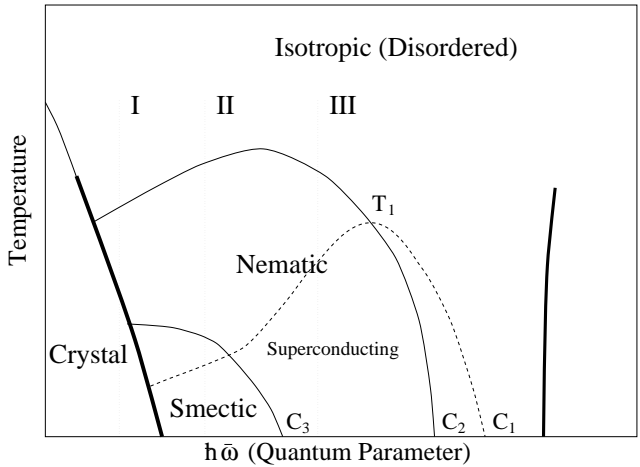}
\vspace{.8cm}
\caption{
Schematic phase diagram of the electronic liquid crystal 
phases of a doped Mott insulator, as discussed in the text.  The
quantum parameter, $\hbar \bar \omega$ depends in a complicated way
on material parameters, as well as the doped hole concentration.  The dotted
lines can be thought of as representing
the temperature evolution of different materials:  
Along I, the insulating stripe crystal ground state melts in
two steps, a sequence of transitions similar to that 
seen\protect \cite{cheong}
in La$_{5/3}$Sr$_{1/3}$NiO$_4$. 
Along II, the simultaneous stripe (smectic) and
superconducting order of the ground state evolve through a sequence 
of transitions in which first the superconducting, then the stripe,
and finally the orientational (nematic) order are lost;
this is reminiscent of the transitions observed
\protect \cite{tranq} in La$_{1.6-x}$Nd$_{0.4}$Sr$_x$CuO$_{4}$.
Along III, the ground state exhibits only superconducting and orientational
order, but the proximity of the quantum critical point, C$_3$, implies that
significant stripe correlations, with dynamics and thermal evolution
governed by this critical point, should be observable at low temperatures;
this is highly reminiscent of the behavior\protect \cite{gabe} of 
La$_{1.86}$Sr$_{0.14}$CuO$_4$ .
}
\label{fig2}
\end{figure}

\section{Quenched Disorder and Stripe Glass Phases}

It was shown by Larkin\cite{larkin} that quenched disorder is relevant 
in CDW systems in dimension $d < 4$, {\it i.e.}, in any physical dimension, true 
long range CDW order simply does not occur in the presence of disorder!  However, 
in high enough dimension and for weak enough disorder, there exists a 
``Bragg glass'' phase,\cite{bragg} which in the present context we refer to as a 
``stripe-glass'', which has power-law density-wave order.  This is a 
distinct state of matter, and hence must be separated from the high 
temperature melted phase by a sharp phase transition.  This phase is 
now more or less established in $d=3$, and it has been 
argued\cite{sudip} that 
in $d=2$, even if an exponentially dilute concentration of free 
dislocations spoils the power-law decay of correlations at very 
long distances, this has very little practical consequence.  For the 
materials of interest to us here, which are either three dimensional 
or quasi-two dimensional, it is safe to conclude that a true stripe 
glass phase exists, with a sharply defined glass transition 
temperature, $T_{g}$. 

Where, as a function of varying material 
parameters, $T_{g}\rightarrow 0$, we expect a quantum critical point.  
However, this quantum critical point will have quite different 
character than those invoked in various theories of high temperature 
superconductivity\cite{quantum}, in that the glass transition is 
disorder driven. Rather, the quantum critical properties will be more 
or less similar to the melting of a ``Wigner glass'', which has been 
invoked to explain the apparent metal-insulator transition observed 
in Si MOSFETs.\cite{sudip} 

Since the charged stripes in doped antiferromagnets are typically, 
anti-phase domain walls in the magnetic order, the 
freezing of the stripe motion at $T_{g}$ opens up the possibility of 
subsequent ({\it i.e.} at still lower temperature) 
ordering of the spins.  As we discussed some time 
ago\cite{physicaC}, spin ordering in a stripe glass will lead to a 
``cluster spin-glass'' phase.   Since spin-glass ordering 
involves a broken symmetry (time reversal), it is easier to detect 
experimentally than stripe-glass ordering, and so more is known about 
the occurrence of this sort of ordering in doped antiferromagnets.
We infer that, in doped antiferromagnets in which cluster 
spin-glass ordering is observed below a spin-glass transition 
temperature $T_{sg}$, there also should be a 
stripe-glass ordering transition with $T_{g} \ge T_{sg}$. 

Finally, it is worth making a couple of general observations 
concerning the effects of quenched disorder.  1)  One of 
the most dramatic features of all glasses is the dramatic slowing 
down of dynamics over a broad range of temperatures as the glass 
transition is approached.  One consequence is that $T_{g}$ is 
always very difficult to determine experimentally (or even to 
determine whether there really is a finite $T_{g}$ at all);  rather, 
there is an apparent $T_{g}$ which depends on the frequency 
at which the system is probed.  Thus, it is to be expected when 
dealing with a glass transition, that the phase diagram, as determined 
by different experimental probes, 
will look rather different, with $T_{g}^{\rm fast} > T_{g}^{\rm slow}$.  
For instance, a glass phase boundary, as determined by neutrons, will be a 
function of the energy resolution, and will extrapolate to the $\mu$-SR 
phase boundary as the energy resolution tends to zero.
2)  It can be proven that disorder eliminates any first order transitions
in two dimensions.\cite{eliminate}  Weak disorder can either 
convert a first order line into a more or less sharp crossover or turn it into a 
continuous transition if there remains a fundamental distinction 
between the two phases.  For instance, in the phase diagram in 
Fig. 2, if the tetracritical point $T_{1}$ were replaced by a bicritical 
point, $B_{1}$, followed by a first order transition to a 
superconducting nematic phase, the first-order line would be replaced by a 
line of continuous phase transitions in the presence of disorder.

\section{Some comments on particular phases observed in the high 
temperature superconductors}

We conclude by making some remarks about a few recent experimental
discoveries in the most studied of doped antiferromagnets, the 
high temperature superconductors, and their relationship to
the theoretical considerations discussed above.  Here, the 
``undoped'' system is a highly-insulating spin-1/2 antiferromagnet, 
which is made conducting and superconducting by the addition of a 
small concentration $x$  of ``doped'' holes.  While much of 
what we discuss is known or presumed to be common to all of the high 
temperature superconductors, we will explicitly refer to experiments 
in the LSCO and YBCO families of materials.
 
\subsection{The insulating ``spin glass''}

For $x$ in the range between about $2\%$ and $5\%$, no long-range order of 
any sort has been observed, and the material is insulating 
at low temperatures.  There is, however, a well defined spin-glass transition.
\cite{budnick}
While the spin-glass phase is generally viewed as a curiosity, of no central 
significance, we have long taken the view\cite{physicaC} that, since this is the 
only ordered phase proximate to the high temperature superconducting 
phase, it should in fact play a central role in our thinking about 
these materials.  We proposed that it is in fact a ``cluster 
spin glass'' phase, in which a frozen random array of stripes produces 
the frustration, so that the spin glass consists of patches of 
locally antiferromagnetically ordered spins with an axis of 
magnetization that varies from patch to patch.  In particular, we 
showed that this proposal naturally accounts for the 
remarkable fact,\cite{keimer} derived from early neutron scattering 
measurements of the spin structure factor, 
that the inverse correlation length, $\kappa(x,T)$, 
obeys the simple composition rule,
\be
\kappa(x,T) = \kappa(0,T) + \kappa(x,0).
\ee

It is clear that there must be a transition at which the stripes freeze into a 
stripe glass.  The spin-glass transtion is more readily detected 
experimentally, since it involves symmetry breaking whereas the 
stripe-glass transition involves only replica-symmetry 
breaking.\cite{bragg,sudip} However, we believe that the stripe glass
is the fundamental phenomenon and that the spin-glass 
transition is more or less parasitic.  Indeed, it is likely that the 
stripe-glass transition temperature is greater than $T_{g}$;  we await 
experimental input on this last issue.

\subsection{The Superconducting stripe glass}

The importance of the spin glass is further substantiated by the old 
observation, which has recently been dramatically 
confirmed by Budnick and collaborators,\cite{budnick} that spin-glass 
and superconducting order in fact coexist in underdoped high 
temperature superconductors!  That static spin order and 
superconducting order can coexist in a single-component electronic 
system is very surprising in terms of conventional paradigms of 
superconductivity.  Of course, the realization that the spin glass is 
actually a stripe glass, which can be viewed as a slightly disordered 
version of the electronic smectic phase discussed above, renders this 
observation a key experimental confirmation of the relevance of 
stripe physics to high temperature superconductivity.  In a sense, 
self-organization into stripes, generates a two-component electronic 
system --- a localized spin component which lives 
between the stripes, and a metallic component which flows along the 
stripes.

In the past few years, neutron scattering studies of {\LNSCO} have 
revealed a still more dramatic and detailed aspect of the coexistence 
of superconducting and magnetic order.  Specifically, Tranquada and 
collaborators\cite{tranq} found a sequence of transitions that depends 
somewhat on Sr concentration. As the temperature decreases there is:  
1)  A structural transition to a low temperature tetragonal (LTT) phase,
2)  A charge-ordering 
transition at which the appearance of well-defined incomensurate elastic 
peaks in the lattice structure factor are driven by charge stripe 
ordering.  3)  A spin-ordering transition, with ordering vector twice 
that of the charge ordering vector.  4)  A superconducting 
transition, with, however, $T_{c}$ reduced relative to that in {\LSCO} 
at the same Sr concentration.  The coexistence of superconducting and stripe 
order was considered surprising, and indeed it has sometimes been attributed 
to sample inhomogeneity. However the evidence 
in favor of coexistence continues to increase.\cite{evidence}
That charge order sets in before spin order confirms that the 
density-wave ordering is ``charge driven'', in the sense defined 
above.  The addition of Nd to the material stabilizes the LTT structure
which allows the oxygen tilting phonon to couple more strongly to 
any charge order - in terms of the schematic phase diagram in Fig. 
2, this reduces the magnitude of the quantum fluctuations of the 
stripes, so the material should be viewed as living farther to the 
left than \LSCO .  Indeed, it is tempting to relate the 
sequence of observed charge transitions to those on a 
trajectory on our phase diagram which passes from the normal state at high temperature, 
through a nematic phase, to a smectic phase, and finally to a 
superconducting smectic phase at low temperatures.  This identification 
is made slightly less than airtight by two subtleties:  1)  It is not 
clear to what extent the structural phase transition to the LTT phase
can be viewed as electronically driven.  2)  The elastic peaks, 
observed in neutron scattering, have a finite width, corresponding to 
a long but finite correlation length for the density-wave order.  
As discussed above, this 
is to be expected in a quasi-two-dimensional system with disorder, 
where any ordered state must be glassy, but it makes the unambiguous
identification of the various phases less secure.

Still more recently, 
neutron scattering experiments on\cite{ky,yamada2} 
underdoped {\LSCO} and 
even\cite{birgeneau}
optimally oxygen-doped {\LCO} (with T$_{c}$ as high as 42K)
have shown that static, fairly long-range 
stripe order and superconductivity coexist.  In these materials, the 
tansition temperatures for spin ordering (which is all that has been 
detected to date) and
superconducting ordering appear to be close to each other, or possibly 
exactly the same.  This demonstrates an intimate 
relation between stripe ordering and superconductivity, and is an 
important new piece of ``theory independent'' evidence for the 
critical role played by stripe order in the mechanism of high 
temperature superconductivity.

{\bf Acknowledgements:}  We would like to aknowledge frequent discussions of
the physics of {\hts} with J.~Tranquada and G.~Aeppli. 
This work was supported at UCLA by the National Science Foundation grant 
number DMR93-12606 and, at Brookhaven,  by the Division of Materials Sciences,
U. S. Department of Energy under contract No. DE-AC02-98CH10886.


\begin{references}

\bibitem{ekl} V.~J.~Emery, S.~A.~Kivelson, and H.-Q.~Lin, 
in {\it Proceedings of the International Conference on the
Physics of Highly correlated Electron Systems, Santa Fe, New Mexico,
September, 1989}, Edited by J.~O.~Willis, J.~D.~Thompson, R.~P.~Guertin, 
and J.~E.~Crow, Physica {\bf B} 163, 306 (1990);
V.~J.~Emery, S.~A.~Kivelson, and H.-Q.~Lin, 
{\prl} {\bf 64}, 475 (1990). 

\bibitem{marder}  M.~Marder, N.~Papanicolau, and G.~C.~Psaltakis, 
{\prb} {\bf 41} 6920 (1990).

\bibitem{manousakis}  S.~Hellberg and E.~Manousakis, {\prl}
{\bf 78}, 4609 (1997).

\bibitem{pryadko}  L.~Pryadko, D.~Hone, and S.~A.~Kivelson, {\prl}
{\bf 80}, 5651 (1998).

\bibitem{topo}  S.~A.~Kivelson and V.~J.~Emery, Synth. Met. {\bf 80}, 
151 (1996).

\bibitem{physicaC}  V.~J.~Emery and S.~A.~Kivelson, Physica {\bf 
C 209}, 597 (1993).

\bibitem{ute}  U.~L\"ow {\it et al.},
Phys. Rev. Lett.  {\bf 72}, 1918  (1994).

\bibitem{science}  M.~Seul and D.~Andelman {\it Science} {\bf 267}, 
476 (1995) and references therein.

\bibitem{larged}  E.~Carlson {\it et al}, {\prb} {\bf 57}, 14704 (1998).

\bibitem{fradkin}  S.~A.~Kivelson, E.~Fradkin, and V.~J.~Emery, {\it 
Nature} {\bf 393}, 550 (1998).

\bibitem{landau}  O.~Zachar, S.~A.~Kivelson, and V.~J.~Emery, {\prb}
{\bf 57}, 1422 (1998).

\bibitem{zaanen}  J.~Zaanen and O.~Gunnarson, {\prb} {\bf 40}, 7391 
(1989).  H.~J.~Schulz, {\prl} {\bf 64}, 1445 (1990).

\bibitem{markku} M.~Salkola, V.~J.~Emery, and S.~A.~Kivelson, {\prl} {\bf 77},
155 (1996).

\bibitem{white}   S.~R.~White and D.~J.~Scalapino {\prl}
 {\bf 80}, 1272 (1998).
For a demonstration that the stripes found here are, in fact, 
insulating see V.~J.~Emery, S.~A.~Kivelson, and O.~Zachar, to be published.

\bibitem{yamada} Yamada {\it et al.} have plotted the spin 
incommensurability $\delta$ as a function of the doped hole concentration
$x_{eff}$ for the LSCO family.\cite{ky} For $x_{eff} < 1/8$, several points 
appear to lie on or close to the line $\delta = x_{eff}$, which corresponds 
to a commensurate doping of one hole per two unit cells along a stripe, which 
would imply insulating stripes. However, the data points actually fall on an 
arc of a circle, which touches the line $\delta = x_{eff}$ near to 
$x_{eff} = 1/8$, the special hole concentration at which T$_c$ is suppressed 
in the LSCO family. 

\bibitem{ky} K. Yamada {\it et al.} {\prb} {\bf 57}, 6165 (1998).

\bibitem{bianconi} A.~Bianconi {\it et al.} in {\it Lattice Effects in
High-T$_c$ Superconductors}, edited by Y.~Bar-Yam. T.~Egami, J.~Mustre-de Leon,
and A.~R.~Bishop (World Scientific, Singapore, 1992), p. 65.

\bibitem{abanov}  Ar.~Abanov {\it et al} {\prb} {\bf 51}, 1023 (1995).

\bibitem{nature}  V.~J.~Emery, S.~A.~Kivelson, {\it Nature} {\bf 434}, 
374 (1995).

\bibitem{ekz}  V.~J.~Emery, S.~A.~Kivelson, and O.~Zachar, {\prb} {\bf 
56}, 6120 (1997).

\bibitem{cheong}  S.~H.~Lee, and S.~W.~Cheong, \prl {\bf 79},
2514 (1997).

\bibitem{gabe}  G.~Aeppli {\it et al}, {\it Science} {\bf 278}, 1432 (1997).

\bibitem{larkin} A. I. Larkin, {\sl Zh. Eksp. Teor Fiz.} {\bf 58}, 
1466 (1970) [{\sl Sov. Phys. JETP},
{\bf 31}, 784 (1970)].

\bibitem{bragg}  T. Giamarchi and P. Le Doussal, {\prb} {\bf 
53}, 15206 (1996); D. Fisher, {\prl} {\bf 78}, 1964 
(1997);  M. J. P. Gingras and D. A. Huse, {\prb} 
{\bf 53}, 15193 (1996).  For a review,
see T. Giamarchi and P. Le Doussal in {\sl Spin Glasses and Random 
Fields}, edited by A. P.  Young
(World Scientific, Singapore, 1997), and references therein.

\bibitem{sudip}  S.~Chakravarty {\it et al}, cond-mat/9805383.

\bibitem{quantum}  See, for example, R.~B.~Laughlin, cond-mat/9709195;
C.~Castellani {\it et al}, Physica C, {\bf 282}, 260 
(1997);  C.~M.~Varma, {\prb} {\bf 55}, 14554 (1997);  S.~Sachdev {\it et 
al}, {\prb} {\bf 51},14874 (1995), and references therein.

\bibitem{eliminate}  Y. Imry and M. Wortis, {\prb} {\bf 19}, 
3580 (1979); M. Aizenman and J. Wehr, 
{\prl}  {\bf 62}, 2503 (1989). A particularly 
transparent physical argument is given by 
A. N. Berker, Physica  {\bf A 194}, 72 (1993). 

\bibitem{budnick}  Ch.~Niedermayer {\it et al} {\prl} {\bf 80}, 3843 
(1998).  See, also, A.~Weidinger {\it et al}, {\prl} {\bf 62}, 102 
(1989) and F.~C.~Chou {\it et al}, \prl {\bf 75}, 2204 (1995).

\bibitem{keimer}  B.~Keimer {\it et al} {\prb} {\bf 46},14034 (1992).
See also S.~M.~Hayden {\it et al}, \prl {\bf 66}, 821 (1991).

\bibitem{tranq}
J.~Tranquada {\it et al.}, {\it Nature} {\bf 375},
561 (1995); J.~M.~Tranquada, {\it Proceedings of the International
Conference on Neutron Scattering} Toronto, Canada (1997), 
Physica {\bf B 241-243}, 745 (1998).

\bibitem{evidence}  J.~E.~Ostenson {\it et al}, {\prb} {\bf 56}, 2820 
(1997).  J.~Tranquada {\it et al}, {\prl} {\bf 78}, 338 (1997).

\bibitem{yamada2} 
T.~Suzuki {\it et al}, {\prb} {\bf 57}, R3229 (1998) and H. Kimura 
{\it et al}, \prb in press.


\bibitem{birgeneau}  Y.~S.~Lee and R.~J.~Birgeneau, private communication.



\end{references}
\end{document}